\documentstyle[12pt]{article}
\begin{document}

\begin{titlepage}
%\hfill 
\vspace{1cm}

\begin{centering}

{\Large \bf Dynamical symmetries in noncommutative theories  }

\vspace{.5cm}
\vspace{1cm}

{\large Ricardo Amorim }

\vspace{0.5cm}

 Instituto de F\'{\i}sica, Universidade Federal
do Rio de Janeiro,\\
Caixa Postal 68528, 21945-970  Rio de Janeiro, Brazil\\[1.5ex]
\vspace{1cm}

\begin{abstract}
In the present work we study dynamical
 space-time symmetries   in noncommutative relativistic theories by using the minimal canonical  extension of the Doplicher, Fredenhagen and Roberts algebra. Our formalism is constructed in an extended space-time with independent degrees of freedom associated with the object of noncommutativity $\theta^{\mu\nu}$. In this framework we consider theories that are invariant under the Poincar\'{e}   group ${\cal P}$ or under its extension ${\cal P}'$, when translations in the extra dimensions are permitted. The Noether's formalism adapted to such extended $x+\theta$ space-time is employed. 
\end{abstract}

\end{centering}

\vspace{1cm}

%\noindent PACS: 03.70.+k, 11.10.Ef, 11.15.-q

\vfill

\noindent{\tt amorim@if.ufrj.br}
\end{titlepage}

\pagebreak

\section{Introduction}
\renewcommand{\theequation}{1.\arabic{equation}}
\setcounter{equation}{0}

\bigskip 
A long time  ago the first work on space-time noncommutativity was written by Snyder\cite{Snyder}. There, the space-time coordinates\footnote{ $A,B=0,1,2,3,4$; $\mu,\nu=0,1,2,3$. The parameter $a$   has dimension of length and  $\hbar=c=1$ .}  $ x^\mu$ have been promoted to operators ${\mathbf x}^{\mu}$ satisfying the  algebra

\begin{eqnarray}
\label{01}
&&[{\mathbf x}^\mu,{\mathbf x}^\nu]= 
i a^2{\mathbf M}^{\mu\nu}\nonumber\\
&&[{\mathbf M}^{\mu\nu},{\mathbf x}^\lambda]=i
({\mathbf x}^{\mu}\eta^{\nu\lambda}-{\mathbf x}^{\nu}\eta^{\mu\lambda})\nonumber\\ 
&&[{\mathbf M}^{\mu\nu},{\mathbf M}^{\alpha\beta}]=i({\mathbf M}^{\mu\beta}\eta^{\nu\alpha}-{\mathbf M}^{\mu\alpha}\eta^{\nu\beta}+{\mathbf M}^{\nu\alpha}\eta^{\mu\beta}-
{\mathbf M}^{\nu\beta}\eta^{\mu\alpha})
\end{eqnarray}

\bigskip 
\noindent which is consistent with the identification ${ \mathbf x}^{\mu}=a\,{\mathbf M}^{4\mu}$,  $M^{AB}$ representing the generators, in some representation, of the group $SO(1,4)$. That work was not very successful in its original motivation, which was the introduction of a natural cutoff for quantum field theories. However, in present times, space-time noncommutativity has been a very studied subject, associated with  strings
\cite{Strings}, noncommutative field theories(NCFT's)\cite{NCFT} and  gravity\cite{QG}, which can be related subjects 
\cite{Hull,SW}.  In NCFT's, usually  the first of relations  (\ref{01}) is  replaced by

\begin{equation}
\label{02}
[{\mathbf x}^\mu,{\mathbf x}^\nu] = i {\mathbf \theta}^{\mu\nu}
\end{equation} 

\bigskip \noindent but in most situations, and contrarily to what occurs in (\ref{01}),
the object of noncommutativity  ${\mathbf \theta}^{\mu\nu}$ is considered as a constant matrix, which implies in the violation of the Lorentz symmetry\cite{NCFT}. A constant $\theta$ is indeed a consequence of the adopted  theory. When strings have their end points on D-branes, in the presence of  a constant antisymmetric tensor field background, this kind of canonical noncommutativity effectively arises. 

It is possible, however, to consider ${\mathbf \theta}^{\mu\nu}$  as an independent operator\cite{Carlson}, resulting in a true Lorentz invariant theory. The results of Ref.\cite{Carlson} have been applied to specific situations\cite{Haghighat,Carone,Ettefasghi} and their consequences have been explored\cite{Morita,Saxell}. 
These works\cite{Carlson}-\cite{Saxell} are based on some contraction of the  algebra (\ref{01}), or equivalently, on the  Doplicher, Fredenhagen and Roberts (DFR) algebra\cite{DFR},
that assumes, besides (\ref{02}), the structure

\begin{eqnarray}
\label{03}
&&[{\mathbf x}^\mu,{\mathbf \theta}^{\alpha\beta}] =0\nonumber\\
&&[{\mathbf \theta}^{\mu\nu},{\mathbf \theta}^{\alpha\beta}]=0
\end{eqnarray}

\bigskip \noindent An important point of the DFR algebra is that the Weyl representation of noncommutative operators obeying (\ref{02}),( \ref{03})
keeps the usual form of the Moyal product, and consequently the form of the usual NCFT's, although the fields have to be considered as depending not only on ${\mathbf x}^\mu$ but also on ${\mathbf \theta}^{\alpha\beta}$.

The DFR algebra has been proposed based in arguments coming from General Relativity and Quantum Mechanics. The authors claim that  very accurate measurements of spacetime localization could transfer to test particles energies sufficient to create a gravitational field that in principle could trap photons. This possibility is related with space-time uncertainty relations that can be derived from (\ref{02}), (\ref{03}) as well as from the quantum conditions

\begin{eqnarray}
\label{04}
&&{\mathbf \theta}_{\mu\nu}{\mathbf \theta}^{\mu\nu} =0\nonumber\\
&&({1\over4}\star{\mathbf \theta}^{\mu\nu}{\mathbf \theta}_{\mu\nu})^2=\lambda_P^8
\end{eqnarray}

\noindent where $\star{\mathbf \theta}_{\mu\nu}={1\over2}\epsilon_{\mu\nu\rho\sigma}{\mathbf \theta}^{\rho\sigma}$and $\lambda_P$ is the Planck length.  These operators are seen as acting on a Hilbert space ${\cal H}$ and this theory implies in extra compact dimensions\cite{DFR}. We observe that the use of the conditions (\ref{04}) in Refs.\cite{Carlson}-\cite{Saxell} would imply in trivial consequences, since  in those works the relevant results  strongly depend in the value of $\theta^2$, actually taken as a mean with some weigh function $W(\theta)$. Of course those authors do not use (\ref{04}), since their motivations are not related with
quantum gravity  but
basically with the construction of a NCFT which keeps Lorentz invariance. This is a fundamental  matter, since there is no experimental evidence to assume Lorentz symmetry  violation\cite{EXP}. 
In the present work we are not considering twisted symmetries\cite{twisted}.

In noncommutative quantum mechanics (NCQM)\cite{Sheikh}-\cite{Rosenbaum}, as in NCFT, a similar framework with constant $\theta$ is usually employed, leading also to the violation of the Lorentz symmetry in the relativistic case or of  symmetry under rotations for  nonrelativistic formulations. In two recent works
\cite{Amorim1,Amorim2} the author has explored some consequences of considering the object of noncommutativity as an independent quantity, respectively as an operator acting in Hilbert space, in the quantum case, or as a phase space coordinate, in the case of classical mechanics. In both situations it was introduced a canonical conjugate momentum for  $\theta$. It has been shown that both theories are related through  the Dirac quantization procedure, once a proper second class constraint structure is postulated.

In the present work we generalize the formalism appearing in \cite{Amorim1}  to the relativistic case, studying how  symmetries can be dynamically
implemented by using the minimal canonical  extension of the DFR algebra. Our formalism, as those of Refs.\cite{Carlson}-\cite{DFR}, is constructed in an extended space-time with independent degrees of freedom associated with the object of noncommutativity $\theta^{\mu\nu}$. In this framework we consider theories that are invariant under the Poincar\'{e}   group ${\cal P}$ and under its extension ${\cal P}'$, when translations in the extra dimensions are also considered, by using the Noether's formalism\cite{Iorio} adapted to such extended space. In Section {\bf 2} we study the algebraic structure of the generalized coordinate operators and their conjugate momenta, and construct the appropriate representations for the generators of ${\cal P}$ and ${\cal P}'$, as well as for the associated Casimir operators. Some possible NCQM actions constructed with those Casimir operators are introduced in Section {\bf3}. In Section {\bf 4} we study the symmetry content of one of those theories by using the Noether's procedure. 
Section {\bf 5} are reserved to some concluding remarks.

\section{Coordinate operators  and their transformations in relativistic NCQM}
\renewcommand{\theequation}{2.\arabic{equation}}
\setcounter{equation}{0}

In the usual formulations of NCQM, interpreted here as relativistic theories, the coordinates ${\mathbf x}^\mu$ and their conjugate momenta ${\mathbf p}_\mu$ are operators acting in a Hilbert space ${\cal H}$  satisfying the 
fundamental commutation relations

\begin{equation}
\label{i1}
[{\mathbf x}^\mu,{\mathbf x}^\nu ] = i{ \theta}^{\mu\nu}
\end{equation}

\begin{equation}
\label{i2}
[{\mathbf x}^\mu,{\mathbf p}_\nu ] = i\delta^{\mu}_\nu
\end{equation}

\begin{equation}
\label{i3}
[{\mathbf p}_\mu,{\mathbf p}_\nu ] = 0
\end{equation}

\noindent and ${\mathbf \theta}^{\mu\nu}$ is considered as a constant matrix.
The ordinary version of  the Lorentz group generator  in the appropriate
representation for acting on space-time coordinates is given by

\begin{equation}
\label{i4}
{ \mathbf l}^{\mu\nu}= { \mathbf x}^\mu{\mathbf p}^\nu-{\mathbf x}^\nu{\mathbf p}^\mu
\end{equation}

\noindent  and does not close in the  Lorentz algebra if (\ref{i1})-(\ref{i3}) are adopted.  Actually

\begin{eqnarray}
\label{i4a}
[{\mathbf l}^{\mu\nu},{\mathbf l}^{\rho\sigma}]&=&i\eta^{\mu\sigma}{\mathbf l}^{\rho\nu}-i\eta^{\nu\sigma}{\mathbf l}^{\rho\mu}-i\eta^{\mu\rho}{\mathbf l}^{\sigma\nu}+i\delta^{\nu\rho}{\mathbf l}^{\sigma\mu}\nonumber\\
&+&i{\mathbf \theta}^{\mu\sigma}{\mathbf p}^{\rho}{\mathbf p}^{\nu}-i{\mathbf \theta}^{\nu\sigma}{\mathbf p}^{\rho}{\mathbf p}^{\mu}-i{\mathbf \theta}^{\mu\rho}{\mathbf p}^{\sigma}
{\mathbf p}^{\nu}+i{\mathbf \theta}^{\nu\rho}{\mathbf p}^{\sigma}{\mathbf p}^{\mu}
\end{eqnarray}

\noindent Furthermore, both sides of (\ref{i1}) would transform in different ways under the action of (\ref{i4}), generating inconsistencies.  An improvement for such a situation can be given by the introduction of the shifted operator\cite{Chaichan,Gamboa}

\begin{equation}
\label{i5}
{\mathbf X}^\mu={\mathbf x}^\mu+{1\over2}{\mathbf\theta}^{\mu\nu}{\mathbf p}_\nu
\end{equation}

\noindent As

\begin{equation}
\label{i6}
[{\mathbf X}^\mu,{\mathbf X}^\nu]=0
\end{equation}

\noindent and 

\begin{equation}
\label{i7}
[{\mathbf X}^\mu,{\mathbf p}_\nu]=i\delta^\mu_\nu\
\end{equation}

\noindent  the operator

\begin{equation}
\label{i8}
{ \mathbf L}^{\mu\nu}= { \mathbf X}^\mu{\mathbf p}^\nu-{\mathbf X}^\nu{\mathbf p}^\mu
\end{equation}

\noindent closes in the Lorentz algebra

\begin{eqnarray}
\label{i9}
[{\mathbf L}^{\mu\nu},{\mathbf L}^{\rho\sigma}]&=&i\eta^{\mu\sigma}{\mathbf L}^{\rho\nu}-i\eta^{\nu\sigma}{\mathbf L}^{\rho\mu}-i\eta^{\mu\rho}{\mathbf L}^{\sigma\nu}+i\delta^{\nu\rho}{\mathbf L}^{\sigma\mu}
\end{eqnarray}

\noindent as can be verified.
By defining the operator

\begin{equation}
\label{i10}
{\mathbf G}_1={1\over2}\omega_{\mu\nu}{\mathbf L}^{\mu\nu}
\end{equation}

\noindent we note that it is possible to dynamically generate infinitesimal transformations on any operator ${\mathbf A}$, following the usual rule
$\delta {\mathbf A}=i[ {\mathbf A}, {\mathbf G}_1]$. For ${\mathbf X}^\mu$,
${\mathbf p}_\mu$ and ${\mathbf L}^{\mu\nu}$ itself we get the expected results

\begin{eqnarray}
\label{i11}
\delta {\mathbf X}^\mu&=&\omega ^\mu_{\,\,\,\,\nu}{\mathbf X}^\nu\nonumber\\ 
\delta{\mathbf p}_\mu&=&\omega _\mu^{\,\,\,\,\nu}{\mathbf p}_\nu\nonumber\\
\delta {\mathbf L}^{\mu\nu}&=&\omega ^\mu_{\,\,\,\,\rho}{\mathbf L}^{\rho\nu}+ \omega ^\nu_{\,\,\,\,\rho}{\mathbf L}^{\mu\rho}
\end{eqnarray}

\noindent but the physical coordinates fail to transform in the appropriate way. As can be seen, the same rule applied on ${\mathbf x}^\mu$ gives

\begin{equation}
\label{i12}
\delta {\mathbf x}^\mu=\omega ^\mu_{\,\,\,\,\nu}
({\mathbf x}^\nu+{1\over2}{\mathbf \theta}^{\rho\nu}{\mathbf p}_\nu)-
{1\over2}{\mathbf \theta}^{\mu\nu}\omega_{\nu\rho}{\mathbf p}^{\rho}
\end{equation}

\noindent which is a consequence of $\theta^{\mu\nu}$ not being transformed. Relation (\ref{i12}) probably will break  Lorentz symmetry in any reasonable  theory. The cure for  these problems can be obtained by considering ${\mathbf \theta}^{\mu\nu}$ as an operator in ${\cal H}$, and introducing its canonical momentum ${\mathbf \pi}_{\mu\nu}$ as well. The price to be paid is that ${\mathbf \theta}^{\mu\nu}$ will have to be associated with extra dimensions, as happens with the formulations appearing in Refs. \cite{Carlson}-\cite{DFR}. 

Let us consider this point with some detail. Following \cite{DFR} we postulate that

\begin{eqnarray}
\label{i13}
&&[{\mathbf x}^\mu,{\mathbf \theta}^{\alpha\beta}] =0\nonumber\\
&&[{\mathbf \theta}^{\mu\nu},{\mathbf \theta}_{\alpha\beta}]=0
\end{eqnarray}

Expressions (\ref{i1}) and (\ref{i13}) form the so called  DFR algebra. 
As we are considering the canonical momentum conjugate to ${\mathbf x}^\mu$, we introduce as well the momenta conjugate to  ${\mathbf \theta}^{\mu\nu}$,
denoted by  ${\mathbf \pi}_{\mu\nu}$ and postulate  that

\begin{eqnarray}
\label{i14}
&&[{\mathbf \theta}^{\mu\nu},{\mathbf \pi}_{\rho\sigma}]=i\delta^{\mu\nu}_{\,\,\,\rho\sigma} \nonumber\\
&&[{\mathbf \pi}_{\mu\nu},{\mathbf \pi}_{\rho\sigma}]=0\nonumber\\
&&[{\mathbf p}_\mu,{\mathbf \theta}^{\rho\sigma}]=0\nonumber\\
&&[{\mathbf p}_\mu,{\mathbf \pi}_{\rho\sigma}]=0
\end{eqnarray}

\noindent where $\delta^{\mu\nu}_{\,\,\,\,\rho\sigma}=\delta^\mu_\rho\delta^\nu_\sigma-\delta^\mu_\sigma\delta^\nu_\rho$.
 Furthermore, we observe that the commutation relation

\begin{equation}
\label{i15}
[{\mathbf x}^\mu,{\mathbf \pi}_{\rho\sigma}]=-{i\over2}\delta^{\mu\nu}_{\,\,\,\rho\sigma}p_\nu
\end{equation}

\noindent is necessary  for algebraic consistency  under Jacobi identities. The set (\ref{i1})-(\ref{i3}),(\ref{i13})-(\ref{i15}) forms the minimal canonical extension of the DFR algebra.

\bigskip

The scheme introduced above permits consistently to adopt\cite{Gracia}

\begin{equation}
\label{i16}
{ \mathbf M}^{\mu\nu}= { \mathbf X}^\mu{\mathbf p}^\nu-{\mathbf X}^\nu{\mathbf p}^\mu-{\mathbf \theta}^{\mu\sigma}{\mathbf \pi}_\sigma^{\,\,\nu}+{\mathbf \theta}^{\nu\sigma}{\mathbf \pi}_\sigma^{\,\,\mu}
\end{equation}

\bigskip\noindent as the generator of the Lorentz group, since it not only closes in the appropriate algebra

\begin{equation}
\label{i17}
[{\mathbf M}^{\mu\nu},{\mathbf M}^{\rho\sigma}]=i\eta^{\mu\sigma}{\mathbf M}^{\rho\nu}-i\eta^{\nu\sigma}{\mathbf M}^{\rho\mu}-i\eta^{\mu\rho}{\mathbf M}^{\sigma\nu}+i\eta^{\nu\rho}{\mathbf M}^{\sigma\mu}
\end{equation}
 
\bigskip\noindent but generates the  expected Lorentz transformations on the Hilbert space operators. Actually, for
$\delta {\mathbf A}=i[ {\mathbf A}, {\mathbf G}_2]$, with $ {\mathbf G}_2={1\over2}\omega_{\mu\nu}{\mathbf M}^{\mu\nu}$, we get

\begin{eqnarray}
\label{i18}
\delta {\mathbf x}^\mu&=&\omega ^\mu_{\,\,\,\,\nu}{\mathbf x}^\nu\nonumber\\ 
\delta {\mathbf X}^\mu&=&\omega ^\mu_{\,\,\,\,\nu}{\mathbf X}^\nu\nonumber\\ 
\delta{\mathbf p}_\mu&=&\omega _\mu^{\,\,\,\,\nu}{\mathbf p}_\nu\nonumber\\
\delta{\mathbf \theta}^{\mu\nu}&=&\omega ^\mu_{\,\,\,\,\rho}{\mathbf \theta}^{\rho\nu}+ \omega ^\nu_{\,\,\,\,\rho}{\mathbf \theta}^{\mu\rho}\nonumber\\
\delta{\mathbf \pi}_{\mu\nu}&=&\omega _\mu^{\,\,\,\,\rho}{\mathbf \pi}_{\rho\nu}+ \omega _\nu^{\,\,\,\,\rho}{\mathbf \pi}_{\mu\rho}\nonumber\\
\delta {\mathbf M}^{\mu\nu}&=&\omega ^\mu_{\,\,\,\,\rho}{\mathbf M}^{\rho\nu}+ \omega ^\nu_{\,\,\,\,\rho}{\mathbf M}^{\mu\rho}
\end{eqnarray}

\bigskip\noindent
which in principle should guarantee the Lorentz invariance of a consistent  theory. We observe that this is only possible because of the introduction of the canonical pair
${\mathbf\theta}^{\mu\nu}$, ${\mathbf \pi}_{\mu\nu}$ as independent  variables. This permits the existence of an object like 
${\mathbf M}^{\mu\nu}$ in (\ref{i16}), which generates the transformations given above  dynamically\cite{Iorio} and not merely by taking in account the index content of the variables, in a pure algebraic way.

By examining the symmetry structure given above, we see that actually the Lorentz generator (\ref{i17}) can be written as the sum of two commuting pieces,
${ \mathbf M}^{\mu\nu}= { \mathbf M}_1^{\mu\nu}+{ \mathbf M}_2^{\mu\nu}$, where
${ \mathbf M}_1^{\mu\nu}= { \mathbf X}^\mu{\mathbf p}^\nu-{\mathbf X}^\nu{\mathbf p}^\mu$ and ${ \mathbf M}^{\mu\nu}_2=
-{\mathbf \theta}^{\mu\sigma}{\mathbf \pi}_\sigma^{\,\,\nu}+{\mathbf \theta}^{\nu\sigma}{\mathbf \pi}_\sigma^{\,\,\mu}$, as in the usual addition of angular momenta. Of course both operators satisfy the Lorentz algebra. It is possible to find convenient representations that reproduce (\ref{i18}). In the sector ${\cal H}_1$ of ${\cal H}={\cal H}_1\otimes{\cal H}_2$ associated with  ${\mathbf X},{\mathbf p}$, it can be used the usual $4\times4$  matrix representation $D_1(\Lambda)=(\Lambda^\mu_{\,\,\,\alpha})$, such that, for instance $ {\mathbf X'}^\mu=\Lambda^\mu_{\,\,\,\nu}{\mathbf X}^\nu$. For the sector of ${\cal H}_2$ associated with  ${\mathbf \theta},{\mathbf \pi}$, it is possible to use the $6\times6$ antisymmetric product representation $D_2(\Lambda)=(\Lambda^{[\mu}_{\,\,\,\alpha}\Lambda^{\nu]}_{\,\,\,\beta})$, such that, for instance, $ {\mathbf \theta'}^{\mu\nu}=\Lambda^{[\mu}_{\,\,\,\alpha}\Lambda^{\nu]}_{\,\,\,\beta}{\mathbf \theta}^{\alpha\beta}$. The complete representation is given by $D=D_1\oplus D_2$.  In the infinitesimal case, $\Lambda^\mu_{\,\,\,\nu}=\delta^\mu_\nu+\omega^\mu_{\,\,\,\nu}$, and (\ref{i18}) are reproduced.
There are four Casimir invariant operators in this context. As can be verified, they are given by  ${\mathbf C_j}_{1}={\mathbf M_j}^{\mu\nu}{ \mathbf M_j}_{\mu\nu}$ and ${ \mathbf C_j}_{2}=\epsilon_{\mu\nu\rho\sigma}{ \mathbf M_j}^{\mu\nu}{ \mathbf M_j}^{\rho\sigma}$, ${\bf j}=1,2$. We note that although the target space has $10=4+6$ dimensions, the symmetry group has only 6 independent parameters
and not the 45 independent parameters of the  Lorentz group in $D=10$.
\bigskip

So it is possible to consistently implement Lorentz symmetry in NCQM following the lines above, once we introduce an appropriate theory, for instance, given by a scalar action.  We know, however, that the  elementary particles are classified according to the eigenvalues of the Casimir operators of the inhomogeneous Lorentz group. In this way it is fundamental to extend this approach to the Poincar\'{e} group ${\cal P}$. By considering the operators presented here, we can in principle consider 
${\mathbf G}_3={1\over2}\omega_{\mu\nu}{\mathbf M}^{\mu\nu}-a^\mu{\mathbf p}_\mu+{1\over2}b_{\mu\nu}{\mathbf \pi}^{\mu\nu}$ as the generator of some   group  ${\cal P}'$, which has the Poincar\'{e} group as a subgroup.  By following the same rule as the one used in the obtainment of (\ref{i18}), with  ${\mathbf G}_2$ replaced by ${\mathbf G}_3$, we arrive now at the set of transformations

\begin{eqnarray}\label{i19}
\delta {\mathbf X}^\mu&=&\omega ^\mu_{\,\,\,\,\nu}{\mathbf X}^\nu+a^\mu\nonumber\\ 
\delta{\mathbf p}_\mu&=&\omega _\mu^{\,\,\,\,\nu}{\mathbf p}_\nu\nonumber\\
\delta{\mathbf \theta}^{\mu\nu}&=&\omega ^\mu_{\,\,\,\,\rho}{\mathbf \theta}^{\rho\nu}+ \omega ^\nu_{\,\,\,\,\rho}{\mathbf \theta}^{\mu\rho}+b^{\mu\nu}\nonumber\\
\delta{\mathbf \pi}_{\mu\nu}&=&\omega _\mu^{\,\,\,\,\rho}{\mathbf \pi}_{\rho\nu}+ \omega _\nu^{\,\,\,\,\rho}{\mathbf \pi}_{\mu\rho}\nonumber\\
\delta {\mathbf M}_1^{\mu\nu}&=&\omega ^\mu_{\,\,\,\,\rho}{\mathbf M}_1^{\rho\nu}+ \omega ^\nu_{\,\,\,\,\rho}{\mathbf M}_1^{\mu\rho}+a^\mu{\mathbf p}^\nu-a^\nu{\mathbf p}^\mu\nonumber\\
\delta {\mathbf M}_2^{\mu\nu}&=&\omega ^\mu_{\,\,\,\,\rho}{\mathbf M}_2^{\rho\nu}+ \omega ^\nu_{\,\,\,\,\rho}{\mathbf M_2}^{\mu\rho}+b^{\mu\rho}{\mathbf \pi}_\rho^{\,\,\,\,\nu}+ b^{\nu\rho}{\mathbf \pi}_{\,\,\,\rho}^{\mu}\nonumber\\
\delta {\mathbf x}^\mu&=&\omega ^\mu_{\,\,\,\,\nu}{\mathbf x}^\nu+a^\mu+{1\over2}b^{\mu\nu}{\mathbf p}_\nu 
\end{eqnarray}

\noindent We observe that there is an unexpected term in the last of relations (\ref{i19}). This is a consequence of the coordinate operator, by (\ref{i5}),
being a nonlinear combination of operators that act on ${\cal H}_1$ and ${\cal H}_2$.

The action of ${\cal P}'$ over the Hilbert space operators is in some sense the action of the 
Poincar\'{e} group with an additional translation on the ${\mathbf \theta}^{\mu\nu}$ sector. All its generators close in an algebra under commutation , so ${\cal P}'$ is a well defined group of transformations. Actually,
the commutation of two  transformations closes in the algebra

\begin{equation}
\label{i19a}
[\delta_2,\delta_1]\,{\mathbf y}=\delta_3\,{\mathbf y}
\end{equation}

\noindent ${\mathbf y}$ representing any one of the operators appearing in (\ref{i19}). The parameters composition rule is given by

\begin{eqnarray}
\label{i19b}
&&\omega^\mu_{3\,\,\,\,\nu}=\omega^\mu_{1\,\,\,\,\alpha}\omega^\alpha_{2\,\,\,\,\nu}-\omega^\mu_{2\,\,\,\,\alpha}\omega^\alpha_{1\,\,\,\,\nu}\nonumber\\
&&a_3^\mu=\omega^\mu_{1\,\,\,\nu}a_2^\nu-\omega^\mu_{2\,\,\,\nu}a_1^\nu\nonumber\\
&&b_3^{\mu\nu}=\omega^\mu_{1\,\,\,\rho}b_2^{\rho\nu}-\omega^\mu_{2\,\,\,\rho}b_1^{\rho\nu}-\omega^\nu_{1\,\,\,\rho}b_2^{\rho\mu}+
\omega^\nu_{2\,\,\,\rho}b_1^{\rho\mu}
\end{eqnarray}

Let us look at the symmetry structure given above. If we consider the operators acting  only on ${\cal H}_1$, we verify that they transform as usual under the Poincar\'{e} group $P$ in $D=4$, whose generators are ${\mathbf p}^\mu$ and ${\mathbf M}_1^{\mu\nu}$. As it is well known, it is formed by the semidirect product between the Lorentz group $L$  in $D=4$ and the translation group $T_4$, and presents two Casimir invariant operators
 ${\mathbf C}_1={\mathbf p}^2$ and ${\mathbf C}_2={\mathbf s}^2$, where 
${\mathbf s}_\mu={1\over2}\epsilon_{\mu\nu\rho\sigma}{\mathbf M}_1^{\nu\rho}{\mathbf p}^\sigma$  is the Pauli-Lubanski vector. If we include in ${\mathbf M}_1$ terms associated with spin, we will keep the usual classification of the elementary particles based on those invariants. 
A representation for $P$ can be given by the $5\times5$ matrix

\begin{equation}\label{P}
D_3(\Lambda,A)=
\pmatrix{\Lambda^\mu_{\,\,\,\nu}&A^\mu\cr
	0&1\cr}
\end{equation}

\noindent acting in the 5-dimensional  vector ${{\mathbf X}^\mu \atopwithdelims ( ) 1}$.

When one considers the operators acting on ${\cal H}_2$, we find a similar structure.  Let us call the corresponding  symmetry group as $G$. It has as generators the operators ${\mathbf \pi}^{\mu\nu}$ and ${\mathbf M}_2^{\mu\nu}$. As one can verify,  ${\mathbf C}_3={\mathbf \pi}^2$ and ${\mathbf C}_4={\mathbf M}_2^{\mu\nu}{\mathbf \pi}_{\mu\nu}$ are the corresponding  Casimir operators. $G$  can be   seen as the semidirect product of the Lorentz group and the translation group $T_6$. A possible representation uses the antisymmetric $6\times6$  representation $D_2(\Lambda)$ already discussed, and is given by the $7\times7$ matrix

\begin{equation}\label{G}
D_4(\Lambda,B)=
\pmatrix{\Lambda^{[\mu}_{\,\,\,\alpha}\Lambda^{\nu]}_\beta&B^{\mu\nu}\cr
	0&1\cr}
\end{equation}

\noindent acting in the 7-dimensional  vector ${\theta^{\mu\nu}\atopwithdelims ( )1}$.
Now we see that the complete group $P$' is just the product of $P$ and $G$. It has a $11\times11$ dimensional representation given by

\begin{equation}\label{P'}
D_5(\Lambda,A, B)=
\pmatrix{\Lambda^\mu_{\,\,\,\nu}&0&A^\mu\cr
0&\Lambda^{[\mu}_{\,\,\,\alpha}\Lambda^{\nu]}_\beta&B^{\mu\nu}\cr
	0&0&1\cr}
\end{equation}

\noindent acting in the 11-dimensional colum vector $\pmatrix{ X^\mu&\cr\theta^{\mu\nu}&\cr 1&}$. A group element needs $6+4+6$ parameters to be determined and $P$' is a subgroup of the full Poincar\'{e}   group $P_{10}$ in $D=10$.  Observe that an element of $P_{10}$ needs 55 parameters to be specified. Here, in the infinitesimal case, when $A$ goes to $a$, $B$ goes to $b$ and $\Lambda^\mu_{\,\,\,\nu}$ goes to $\delta^\mu_\nu+\omega^\mu_{\,\,\,\nu}$, the transformations (\ref{i19}) are obtained from the action of (\ref{P'}) defined above. Naturally ${\mathbf C}_1$, ${\mathbf C}_2$, ${\mathbf C}_3$ and ${\mathbf C}_4$ are the Casimir operators of $P$'.

So far we have been considering a possible algebraic structure among operators in ${\cal H}$ and  possible sets of transformations for these operators. The choice of an specific theory, however, will give  the mandatory 
criterion for selecting among these sets of transformations, the one that gives the dynamical symmetries of the action. If the considered theory is not invariant under the $\theta$ translations, but it is by Lorentz transformations and $x$ translations, the set of the symmetry transformations on the generalized coordinates will be given by (\ref{i19}) but effectively considering 
 $b_{\mu\nu}$ as vanishing, which implies that $P'$, with this condition, is  dynamically contracted to the Poincar\'{e} group.  Observe, however, that   ${\mathbf \pi}_{\mu\nu}$ will be yet a relevant operator, since ${\mathbf M}_{\mu\nu}$ depends on it in the representation here adopted. 
An important point related with the dynamical action of ${\cal P}$ is that it conserves the quantum conditions (\ref{04}).

In the next section we are going to consider some  points concerning some actions which give  models for a relativistic NCQM  in order to derive their equations of motion and display their symmetry content in Section {\bf 4}.

\section{Actions}
\renewcommand{\theequation}{3.\arabic{equation}}
\setcounter{equation}{0}

As discussed in the previous section, in NCQM the physical coordinates do not commute and their eigenvectors can not be used to form a basis in ${\cal H}={\cal H}_1+{\cal H}_2$.
This does not occur with the shifted coordinate operator ${\mathbf X}^\mu  $ due to (\ref{i6}) and (\ref{i13}). Consequently their eigenvectors can be used in the construction of such a basis. Generalizing what has been done in \cite{Amorim1},  it is possible to introduce a 
 coordinate basis $|X',\theta'>=|X'>\otimes\,|\theta'>$ in such a way that

\begin{eqnarray}\label{b1}
 {\mathbf X}^\mu|X',\theta'>&=&{X'}^\mu|X',\theta'>\nonumber\\
{\mathbf\theta}^{\mu\nu} |X',\theta'> &=&{\theta'}^{\mu\nu}|X',\theta'>
\end{eqnarray}

\bigskip\noindent satisfying usual orthonormality and completeness relations. In this basis

\begin{equation}
\label{b2}
< { X}',{ \theta}'|{\mathbf p}_\mu|{ X}",{ \theta}">= -i{\frac{\partial}{\partial X'^\mu}}\delta^{4} (X'-X")\delta^{6}({\theta}'-{\theta}")
\end{equation}

\bigskip\noindent             and

\begin{equation}
\label{b3}
< { X}',{ \theta}'|{\mathbf \pi}_{\mu\nu}|{ X}",{ \theta}">= -i\delta^{4} (X'-X"){\frac{\partial}{\partial \theta'^{\mu\nu}}}\delta^{6}({\theta}'-{\theta}")
\end{equation}

\bigskip\noindent implying that both momenta acquire a derivative realization.

\bigskip

A physical state $|\phi>$,  in the coordinate basis defined above, will be represented by the wave function $\phi(X',\theta')=<X',\theta'|\phi>$
satisfying some wave equation that we assume that can be derived from an action, through a variational principle. 
As it is well known,  a  direct route for constructing  an ordinary relativistic free quantum theory is to impose that the physical states are annihilated by the mass shell condition

\begin{equation}
\label{b4}
({\mathbf p}^2+m^2)|\phi>=0
\end{equation}

\noindent  constructed with the Casimir operator ${\mathbf C}_1={\mathbf p}^2$. In the coordinate representation, this gives the Klein-Gordon equation. 
The same result is obtained from the quantization of the classical relativistic particle, whose action is invariant under reparametrization\cite{Dirac}. There the generator of
the reparametrization symmetry is the constraint $({\mathbf p}^2+m^2)\approx0$. Condition (\ref{b4}) is then interpreted as the one that selects the physical states, that must be invariant under gauge (reparametrization) transformations.
In the noncommutative case, besides (\ref{b4}), it is reasonable to assume as  well that the second condition

\begin{equation}
\label{b5}
({\mathbf \pi}^2+\Delta)|\phi>=0
\end{equation}

\noindent must be imposed on the physical states, since it is also an invariant, and it is not affected by the evolution generated by (\ref{b4}). It can be shown that in the underlying classical theory\cite{NEXT}, this condition is also associated with a first class constraint, which generates gauge transformations, and so (\ref{b5}) can also be seen as selecting gauge invariant states.
In (\ref{b5}), $\Delta$ is some constant with dimension of $M^4$, whose sign and value depend if ${\mathbf \pi}$ is space-like, time-like or null. Both equations
permit to construct a generalized plane wave solution  $\phi(X',\theta')\equiv<X',\theta'|\phi>\sim exp(ik_\mu{X'}^\mu+{i\over2}K_{\mu\nu}{\theta'}^{\mu\nu})$,
where $k^2+m^2=0$ and $K^2+\Delta=0$. In coordinate representation (\ref{b1})-(\ref{b3}), (\ref{b4}) gives just the Klein-Gordon equation

\begin{equation}
\label{b6}
(\Box_X-m^2)\phi(X',\theta')=0
\end{equation}

\noindent while (\ref{b5}) gives the subsidiary equation

\begin{equation}
\label{b7}
(\Box_{\theta}-\Delta)\phi(X',\theta')=0
\end{equation}

\noindent where $\Box_X= \partial^\mu\partial_\mu  $, with $\partial_\mu={{\partial}\over{\partial {X'}^\mu}}$. Also  $\Box_{\theta}={1\over2}\partial^{\mu\nu}\partial_{\mu\nu}$,
with $\partial_{\mu\nu}={{\partial\,\,\,}\over{\partial {\theta'}^{\mu\nu}}}$. Both equations can be derived, at least in the case (i) defined in what follows, from the action

\begin{equation}
\label{b8a}
S=\int d^{4}\,X'\,d^{6}\theta'\,\Omega(\theta')\,\{ {1\over2}(\,\partial^\mu\phi\partial_\mu\phi+m^2\,\phi^2)\, -\Lambda( \Box_\theta -\Delta)\phi\} 
\end{equation}

\noindent  In (\ref{b8}) $\Lambda$ is a Lagrange multiplier necessary to impose
condition (\ref{b7}). $\Omega(\theta')$ can be seen as (i) as a simple constant $\theta_0^{-6}$ to keep the usual dimensions of the  fields as $S$ must be adimensional in natural units, as (ii) an even weight function as the one appearing the Refs. \cite{Carlson}-\cite{Saxell} used to make the connection between the formalism in $D=4+6$ and the usual one in $D=4$ after the integration in $\theta'$, or (iii), a distribution used to impose further conditions as those appearing in (\ref{04}) and adopted in  \cite{DFR}.

A simpler model not involving Lagrangian multipliers, but two of the Casimir operators of ${\cal P}'$, ${\mathbf C}_1={\mathbf p}^2$ and ${\mathbf C}_3={\mathbf \pi}^2$, is given by

\begin{equation}
\label{b8}
S=\int d^{4}\,X'\,d^{6}\theta'\,\Omega(\theta')\, {1\over2}\Big\{\,\partial^\mu\phi\partial_\mu\phi + \,{{\lambda^2}\over4}\,\partial^{\mu\nu}\phi\partial_{\mu\nu}\phi   +m^2\,\phi^2\Big\}\, 
\end{equation}

\noindent where $\lambda$ is a parameter with dimension of length, as the Planck length, which has to be introduced by dimensional reasons. If it goes to zero one essentially obtains the Klein-Gordon action. By borrowing the Dirac matrices $\Gamma_A$, $A=0,1,....,9$, written for space-time $D=10$, and identifying the tensor indices with the six last values of $A$, it is also possible to construct the 'square root' of the equation of motion derived from(\ref{b8}), obtaining a generalized Dirac theory involving spin and noncommutativity\cite{NEXT}.

\bigskip
In the next section we will consider the equations of motion and the Noether's theorem derived for general theories defined in $x+\theta$ space, and specifically  for the action (\ref{b8}), considering $\Omega(\theta)$ as a well behaved function.

\section{Equations of motion and Noether's theorem}
\renewcommand{\theequation}{4.\arabic{equation}}
\setcounter{equation}{0}

Let us consider the action

\begin{equation}
\label{c1}
S=\int_R d^{4}\,x\,d^{6}\theta \,\Omega(\theta)\,{\cal L}(\phi^i,\partial_\mu\phi^i,\partial_{\mu\nu}\phi^i,x, \theta)
\end{equation}

\noindent depending on a set of fields $\phi^i$, their first derivatives with respect to ${x}^\mu$ and ${\theta}^{\mu\nu}$ and the
coordinates ${x}^\mu$ and ${\theta}^{\mu\nu}$ themselves. From now on we will use $x$ in place of $X'$ and $\theta$ in place of $\theta'$ in order to simplify the notation. Naturally the fields $\phi^i$ can be functions of ${x}^\mu$ and ${\theta}^{\mu\nu}$. The index $i$ permits to treat $\phi$ in a general way. In (\ref{c1}) we consider, as in (\ref{b8}), the integration element modified by the introduction of $\Omega(\theta)$.

By assuming that $S$ is stationary for an arbitrary variation  $\delta\phi^i$  vanishing on the boundary $\partial R$ of the
 region of integration $R$, we get the Euler-Lagrange equation

\begin{equation}
\label{c2}
\Omega\big({{\partial{\cal L}}\over{\partial\phi^i}}-\partial_\mu{{\partial{\cal L}}\over{\partial\partial_\mu\phi^i}}\big)-\partial_{\mu\nu}\big(\Omega{{\partial{\cal L}}\over{\partial\partial_{\mu\nu}\phi^i}}\big)=0
\end{equation}

\noindent  Now consider variations  $\delta x^\mu$, $\delta\theta^{\mu\nu}$ of the generalized coordinates and $\delta\phi^i$ of the fields such that the integrand transforms as a total divergence in the $x+\theta$ space,   $\delta(\Omega\,{\cal L})=\partial_\mu(\Omega\,S^\mu)+\partial_{\mu\nu}(\Omega\,S^{\mu\nu})$. Then the Noether's theorem assures that, on shell, or when (\ref{c2}) is satisfied, there is a  conserved current $(j^\mu,j^{\mu\nu})$ defined by

\begin{eqnarray}
\label{c3}
j^\mu&=&{{\partial{\cal L}}\over{\partial\partial_\mu\phi^i}}\delta\phi^i+{\cal L}\delta x^\mu\nonumber\\
j^{\mu\nu}&=&{{\partial{\cal L}}\over{\partial\partial_{\mu\nu}\phi^i}}\delta\phi^i+{\cal L}\delta\theta^{\mu\nu}
\end{eqnarray}

\noindent such that

\begin{equation}
\label{c4}
\Xi= \partial_\mu(\Omega\, j^\mu)+\partial_{\mu\nu}(\Omega\,j^{\mu\nu})
\end{equation}

\noindent vanishes.
The corresponding charge

\begin{equation}
\label{c5}
Q=  \int d^{3}\,x\,d^{6}\theta\,\Omega(\theta)\,j^0 
\end{equation}

\noindent is independent of the 'time' ${x}^0$. Conversely, if exists a conserved current like (\ref{c3}), the action (\ref{c1}) is invariant under the corresponding symmetry transformations. This is just a trivial extension of the usual version of Noether's theorem\cite{Iorio} in order to include $\theta^{\mu\nu}$ as independent coordinates, as well as a modified integration element due to the presence of $\Omega(\theta)$.  We comment that $\Omega$ has not been included in the current definition (\ref{c3}) because it is seen as part of the element of integration,  but it is present in (\ref{c4}), which is the relevant divergence. It is present also in the charge (\ref{c5}) since the charge is an integrated quantity.

\bigskip

Let us  apply (\ref{c1}-\ref{c4}) to the simple model given by (\ref{b8}). The Lagrange equation reads

\begin{eqnarray}
\label{c6}
{{\delta S}\over{\delta\phi}}&=&-\,\Omega\,(\Box - m^2)\phi-{{\lambda^2}\over2}\partial_{\mu\nu}(\Omega\,\partial^{\mu\nu}\phi)\nonumber\\
&=&\,\,\,0
\end{eqnarray}

\noindent while (\ref{c4}) can be written as

\begin{eqnarray}
\label{c61}
\Xi&=&\partial_\mu\Big\{\Omega\,\partial^\mu\phi\,\delta\phi+{{\Omega}\over{2}}\Big(\partial_\alpha\phi\partial^\alpha\phi+{{\lambda^2}\over4}\partial_{\alpha\beta}\phi\partial^{\alpha\beta}\phi+m^2\phi^2\Big)\delta x^\mu\Big\}\\
&+&\partial_{\mu\nu}\Big\{\Omega\,\lambda^2\partial^{\mu\nu}\phi\,\delta\phi+{{\Omega}\over{2}}\Big(\partial_\alpha\phi\partial^\alpha\phi+{{\lambda^2}\over4}\partial_{\alpha\beta}\phi\partial^{\alpha\beta}\phi+m^2\phi^2\Big)\delta\theta^{\mu\nu}\Big\}\nonumber
\end{eqnarray}

Before using (\ref{c61}) we observe that the transformation

\begin{equation}
\label{i19c}
\delta \phi=-(a^\mu+\omega^\mu_{\,\,\,\nu}x^\nu)\,\partial_\mu\phi-{1\over2}(b^{\mu\nu}+2\omega^\mu_{\,\,\,\rho}\theta^{\rho\nu})\,\partial_{\mu\nu}\phi
\end{equation}

\noindent closes in an algebra, as in (\ref{i19a}), with the same composition rule defined in (\ref{i19b}). The above equation defines  how a scalar field transforms in the $x+\theta$ space under the action of ${\cal P}'$.

Let us now consider (i)  a rigid $x$-translation, given by

\begin{eqnarray}
\label{c62}
\delta_a x^\mu&=&a^\mu\nonumber\\
  \delta_a\theta^{\mu\nu}&=&0\nonumber\\
\delta_a\phi&=&-a^\mu\partial_\mu\phi
\end{eqnarray}

\noindent where $a^\mu$ is constant.
We see from (\ref{c61}) and (\ref{c62}) that

\begin{equation}
\label{c63}
\Xi_a=a^\mu\partial_\mu\phi\,\,{{\delta S}\over{\delta\phi}}
\end{equation}

\noindent vanishing on shell, when (\ref{c6}) is valid.

For (ii) a rigid $\theta$-translation, given by

\begin{eqnarray}
\label{c64}
\delta_b x^\mu&=&0\nonumber\\
  \delta_b\theta^{\mu\nu}&=&b^{\mu\nu}\nonumber\\
\delta_b\phi&=&-{1\over2}b^{\mu\nu}\partial_{\mu\nu}\phi
\end{eqnarray}

\noindent where $b^{\mu\nu}$ is constant, we get

\begin{equation}
\label{c65}
\Xi_b={1\over2}b^{\mu\nu}\Big(\partial_{\mu\nu}\phi\,\,{{\delta S}\over{\delta\phi}}+{\cal L}\partial_{\mu\nu}\Omega\Big)
\end{equation}

The first term on the right vanishes on shell but the second one depends on the form of $\Omega$. Later we will comment this point.
At last let us consider (iii) a Lorentz transformation, given by

\begin{eqnarray}
\label{c66}
\delta_\omega x^\mu&=&\omega^{\mu}_{\,\,\,\nu} x^\nu\nonumber\\
\delta_\omega\theta^{\mu\nu}&=&\omega^{\mu}_{\,\,\,\rho}\theta^{\rho\nu}+\omega^{\nu}_\rho\theta^{\mu\rho}\nonumber\\
\delta_\omega\phi&=&-(\omega^\mu_{\,\,\,\nu} x^\nu\partial_\mu+ \omega^{\mu}_{\,\,\,\rho}\theta^{\rho\nu}\partial_{\mu\nu})             \phi
\end{eqnarray}

\noindent with constant and antisymmetric $\omega^\mu_{\,\,\,\nu}$. We get

\begin{equation}
\label{c67}
\Xi_\omega=\omega^{\mu}_{\,\,\,\nu}\,{{\delta S}\over{\delta\phi}}(x^\nu\partial_\mu\phi+\theta^{\nu\rho}\partial_{\mu\rho}\phi)+{\cal L}\partial_{\mu\nu}\Omega\omega^\mu_{\,\,\,\alpha}\theta^{\alpha\nu}
\end{equation}

\noindent The first term in the above expression vanishes on shell and the second one also vanishes if $\Omega$ is a scalar under Lorentz transformations and depends only on $\theta$.

 So we see that the theory described by action (\ref{b8}) is invariant under the Poincar\'{e}  group ${\cal P}$ if $\Omega$ is a scalar depending only on $\theta$ but is not a constant. If $\Omega$ is a constant, it is not only invariant under ${\cal P}$, but also under ${\cal P'}$, as discussed in Section {\bf 2}.
 Of course, in a complete theory where other contributions for the total action would be present , the symmetry under $\theta$ translations could be broken by different reasons, as  in what follows, in the case of the noncommutative $U(1)$ gauge theory. In this situation $\cal P$ could be the symmetry group of the complete theory even considering $\Omega(\theta)$ as a constant.

\section{Conclusions}
 To close this work, we observe that it has been  possible to consistently treat the object of noncommutativity $\theta^{\mu\nu}$ and its canonical conjugate momentum  as  Hilbert space operators, implementing  the minimal canonical  extension of the DFR algebra.  In this framework we could 
construct a generalized Lorentz generator as well as generalized translation operators, which permitted to
consider theories that are invariant under the action of the Poincar\'{e}   group ${\cal P}$ and under its extension ${\cal P}'$, when the translations in the extra dimensions are also taken in account. Representations for both $P$ and $P$'have been given and the Casimir operators of such symmetry groups have been displayed. We also have  considered several possible actions that could be the starting point for NCFT's. Their symmetry content has been explored in a dynamical way,
by using  the Noether's formalism in such extended space. Contrary to what occurs in the usual treatments, we could prove the invariance of those noncommutative theories under the action of ${\cal P}$ or ${\cal P}'$ in a dynamical way.

The formulation here proposed takes in account noncommutativity without destroying the symmetry content of the corresponding commutative theories. We expect that the new features associated with the objects of noncommutativity will be relevant at high energy scales. Even if excited states in the Hilbert space sector associated with noncommutativity are not assessable,  ground state  effects could  in principle be detectable.

\end{document}